\documentclass[twocolumn,amssymb, amsmath,
tightenlines, aps, prl,showpacs,preprintnumbers,floatfix]{revtex4} 

\usepackage{graphicx}
\usepackage{dcolumn}
\usepackage{bm}

\begin{document}


\title{Testing the stability of fundamental constants \\ 
with the $^{199}$Hg$^{+}$ single-ion optical clock}

\author{S. Bize}
    \altaffiliation[Present address: ]
    {BNM-SYRTE, Observatoire de Paris, France}
    \email{sebastien.bize@obspm.fr}
\author{S.A. Diddams}
\author{U. Tanaka}
\author{C.E. Tanner}
    \altaffiliation[Permanent address: ]
    {Department of Physics, University of Notre Dame, also supported by DOE and NSF.}
\author{W.H. Oskay}
\author{R.E. Drullinger}
\author{T.E. Parker}
\author{T.P. Heavner}
\author{S.R. Jefferts}
\author{L. Hollberg}
\author{W.M. Itano}
\author{D.J. Wineland}
\author{J.C. Bergquist}
    \email{berky@boulder.nist.gov}
\affiliation{Time and Frequency Division, National Institute of Standards 
and Technology, 325 Broadway, Boulder, CO 80305, USA} 

\date{\today}

\begin{abstract}
Over a two-year duration, we have compared the frequency of the
$^{199}$Hg$^{+}$ $5d^{10}6s~^{2}S_{1/2} (F=0)\longleftrightarrow
5d^{9}6s^2~^{2}D_{5/2} (F=2)$ electric-quadrupole transition at
$282$~nm with the frequency of the ground-state hyperfine
splitting in neutral $^{133}$Cs. These measurements show that any
fractional time variation of the ratio
$\nu_{\mathrm{Cs}}/\nu_{\mathrm{Hg}}$ between the two frequencies
is smaller than $\pm 7\times10^{-15}$yr$^{-1}$ ($1\sigma$
uncertainty). According to recent atomic structure calculations,
this sets an upper limit to a possible fractional time variation
of $g_{\mathrm{Cs}}(m_e/m_p)\alpha^{6.0}$ at the same level.
\end{abstract}
\maketitle

The development of string theory models aiming at a unified
description of gravity and quantum mechanics has renewed interest
for improved experimental tests of Einstein's Equivalence
Principle (EEP). Indeed, a common feature of these models is that
they allow for, or even predict, violations of EEP
\cite{Damour2002}. These include violation of the universality of
free-fall as well as variation of fundamental constants with time
and space. Interestingly, a recent analysis of the spectrum of
quasars \cite{Webb2001} suggests that the fine-structure constant
$\alpha$ may have changed over the cosmological timescale
($10^{10}$~yr). The Oklo reactor analysis
\cite{Shlyakhter1976,Damour1996} on the other hand puts a
stringent limit to possible variation of $\alpha$ on the
geological timescale ($10^9$~yr). 
Owing to their high accuracy, comparisons between atomic frequency
standards based on different atomic species and/or types of
transitions provide one of the best ways to perform laboratory
tests of the stability of fundamental constants. Present and
future efforts to improve atomic frequency standards, in both the
optical and the microwave domains, will improve these tests,
leading to significant constraints on theoretical work aimed at a
unified theory.

In this letter we describe frequency comparisons conducted over a
two-year period between a $^{199}$Hg$^{+}$ single-ion optical
clock and a $^{133}$Cs fountain atomic clock that set a new
stringent limit to a possible variation of fundamental constants.
The theoretical background for such a test is given first. We then
describe the experiment and conclude with the results of the test.

In the experiment, the frequency $\nu_{\mathrm{Hg}}$ of the
$^{199}$Hg$^{+}$ $5d^{10}6s~^{2}S_{1/2} (F=0)\longleftrightarrow
5d^{9}6s^2~^{2}D_{5/2} (F=2,m_{F}=0)$ electric-quadrupole
transition at $\lambda = 282$~nm is compared to the frequency
$\nu_{\mathrm{Cs}}$ of the ground-state hyperfine transition
$6S_{1/2}(F=3,m_{F}=0)\longleftrightarrow 6S_{1/2}(F=4,m_{F}=0)$
in neutral $^{133}$Cs. Including relativistic and many-body
effects, $\nu_{\mathrm{Hg}}$ can be expressed as
$\nu_{\mathrm{Hg}}\simeq R_{y}~F_{\mathrm{Hg}}(\alpha)$, where
$R_{y}=R_{\infty} c$ is the Rydberg constant expressed as a
frequency, and $F_{\mathrm{Hg}}(\alpha)$ is a dimensionless
function of the fine-structure constant
$\alpha=e^{2}/4\pi\varepsilon_{0}\hbar c$. Similarly, the
hyperfine frequency of cesium can be approximated by
$\nu_{\mathrm{Cs}}\simeq g_{\mathrm{Cs}}(m_{e}/m_{p})\alpha^{2}
R_{y}~F_{\mathrm{Cs}}(\alpha)$, where $g_{\mathrm{Cs}}$ is the
gyro-magnetic ratio of the $^{133}$Cs nucleus and $m_{e}/m_{p}$
the electron-to-proton mass ratio. $F_{\mathrm{Hg}}(\alpha)$ and
$F_{\mathrm{Cs}}(\alpha)$ are calculated in
\cite{Prestage1995,Dzuba1999a}. With numerical values from
\cite{Dzuba1999a}, we find
\begin{eqnarray}
    \alpha~\frac{\partial}{\partial\alpha}\ln
    F_{\mathrm{Hg}}(\alpha)\simeq -3.2
    \\ \alpha~\frac{\partial}{\partial\alpha}\ln F_{\mathrm{Cs}}(\alpha)\simeq +0.8.
     \label{eq:sensitivities}
\end{eqnarray}
Therefore, sequential measurements of the ratio
$\nu_{\mathrm{Cs}}/\nu_{\mathrm{Hg}}$ actually test the stability
of the product of fundamental constants $g_{\mathrm{Cs}}(m_e/m_p)
\alpha^{6.0}$. Note that the sensitivity to variation of the
individual constants needs to be known to only 1-10~$\%$ in order
to adequately describe how this comparison
constrains possible variations of these fundamental constants. It
is therefore justified to omit higher-order terms in the
expression of the hyperfine splitting of $^{133}$Cs, such as
finite nuclear size. The high sensitivity of the Hg$^+$ vs Cs
frequency comparison to a change of $\alpha$ arises from the large
relativistic effects encountered in heavy atoms combined with the
negative sign of the relativistic effects in $^{199}$Hg$^+$. These
factors make the $^{199}$Hg$^{+}$ $^2
S_{1/2}\longleftrightarrow~^2 D_{5/2}$ optical transition one of
the best choices for a search of a variation of $\alpha$
\cite{Dzuba1999a}.
\begin{figure}
\includegraphics[width=6cm]{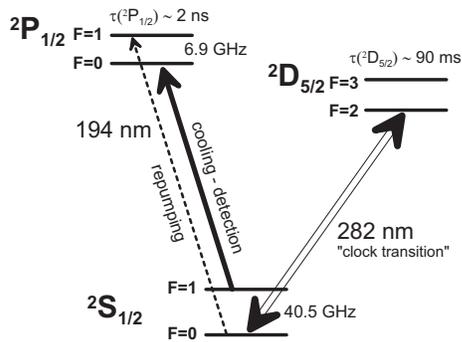}
\caption{\label{fig:energylevels} Partial level scheme of
$^{199}$Hg$^{+}$. The 194~nm $^2S_{1/2}(F=1)\longleftrightarrow
~^2P_{1/2}(F=0)$ transition is used for Doppler cooling, state
preparation and detection. The 282~nm electric-quadrupole
transition from the ground state $^{2}S_{1/2} (F=0)$ to the
metastable $^{2}D_{5/2} (F=2,m_{F}=0)$ state provides the
reference for the optical clock frequency.}
\end{figure}

The $^{199}$Hg$^{+}$ single-ion optical frequency standard has been
described previously \cite{Young1999,Rafac2000} and only the main
features are outlined here.  A single $^{199}$Hg$^{+}$ ion is stored
in a radio-frequency (rf) spherical Paul trap held at a cryogenic
temperature ($4.2$~K).  It is laser-cooled to near the $1.7$~mK
Doppler limit using the strongly allowed
$^2S_{1/2}(F=1)\longleftrightarrow~^2P_{1/2}(F=0)$ transition at
194~nm, as shown in Fig.  \ref{fig:energylevels}.  After a cooling and
state preparation phase ($\sim 16$~ms), the ion is left in the
$^2S_{1/2}(F=0)$ lower state of the clock transition.  The cooling
laser is then switched off and the probe laser light at
$282~\mathrm{nm}$ is directed onto the ion for a typical duration of
$T=50$~ms.  The cooling laser is turned on again to determine the
ion's internal state using the technique of electron shelving
\cite{Dehmelt1975,Bergquist1987}.  The $194~\mathrm{nm}$ fluorescence
photons emitted by the ion are counted for $15$~ms.  The absence of
scattered photons indicates that the ion has been excited to the
$^2D_{5/2}(F=2)$ state by the $282~\mathrm{nm}$ probe laser. 
Similarly, a typical count rate of $\sim 6000$~s$^{-1}$ from the ion
indicates that it has remained in the $^2S_{1/2}(F=0)$ state after the
probe period.  When the ion is detected in the $^2D_{5/2}$ excited
state, the $194~\mathrm{nm}$ is left on until scattered photons are
detected again, indicating that the ion has spontaneously decayed to
the $^2S_{1/2}$ state.  A new interrogation cycle is then started.

\begin{figure}[b]
\includegraphics[width=6cm]{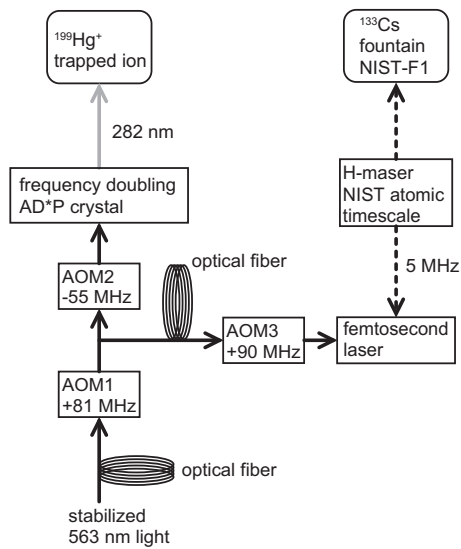}
\caption{\label{fig:measurementscheme} Experimental setup for the
absolute measurement of the frequency of the $^{199}$Hg$^+$
optical clock in terms of the SI second defined by the ground-state
hyperfine splitting of $^{133}$Cs.}
\end{figure}

The 282~nm radiation used to probe the clock transition is obtained by
frequency-doubling 563~nm light from a dye laser in a deuterated
ammonium dihydrogen phosphate (AD$^{\star}$P) crystal, as shown in
Fig.~\ref{fig:measurementscheme}.  The light from the dye laser is
stabilized to a resonance of a stable high-finesse ($\mathcal{F}\sim
200~000$) Fabry-P\'erot cavity.  Two acousto-optic modulators
(AOM1,AOM2) driven by precision rf sources
shift and fine-tune the frequency of the 563~nm radiation so as to
match the frequency of the 282~nm radiation to the $S-D$ resonance
frequency.  By compensating the predictable linear drift of the stable
reference cavity ($\sim 500$~mHz~s$^{-1}$) using AOM1, we realize an
interrogation oscillator with a fractional frequency instability of
$3~\times 10^{-16}$ between 1~s and 10~s, corresponding to a linewidth
of $640$~mHz at 282~nm \cite{Young1999}.  For a typical probe time
$T=50$~ms, which is shorter than the $90$~ms spontaneous decay time of
the $^2 D_{5/2}$ state, the observed linewidth is Fourier-transform
limited with a full width at half maximum of $\sim 16$~Hz, much larger
than the frequency fluctuations of the probe laser during the probe
pulse time.  Under these conditions, the fluctuations of the measured
transition probability, and hence the frequency stability of the
Hg$^{+}$ frequency standard, are limited mainly by atomic quantum
projection noise \cite{Itano1993}.

In our trap, the radial and axial secular frequencies are 1.1~MHz
and 2.2~MHz respectively.  Both frequencies are much higher than
the recoil frequency $\nu=h/2m\lambda^2\approx 12.6~\mathrm{kHz}$, where
$m$ is the ion mass and $h$ is the Planck constant. The
observed spectrum therefore consists of a central feature at the
$^2 S_{1/2}(F=0)\longleftrightarrow~^2 D_{5/2}(F=2,m_{F}=0)$
transition resonance frequency, which is free of the recoil
frequency shift, together with vibrational sidebands at integer
multiples of the trap secular frequencies \cite{Bergquist1987}.
The spectrum also exhibits features corresponding to the carrier
and vibrational sidebands of the $^2
S_{1/2}(F=0)\longleftrightarrow~^2 D_{5/2}(F=2,m_{F}=\pm 1,\pm 2)$
``Zeeman components.'' Typically, the frequency difference between
the clock transition and the first Zeeman components is $\sim
6$~MHz, which corresponds to a quantization magnetic field
$B\approx 0.3$~mT.

In order to lock the frequency of the probe laser to the atomic
resonance, a square-wave frequency modulation is applied to the
563~nm beam using AOM2 so as to probe the carrier of the clock
transition alternately on each side of the resonance. Typically,
24 measurements are averaged on each side of the resonance to find
the transition probability. The difference between these two
transition probabilities reveals the detuning between the center
frequency of the probe laser and the center of the atomic
resonance. This error signal is used in a digital servo loop that
steers the average frequency of the probe laser to the center of
the atomic resonance, applying frequency corrections to the
synthesizer driving AOM1. The time constant of this servo loop is
on the order of $\tau_{\mathrm{loop}}\sim 15$~s. The analysis of
the frequency corrections indicates that for
$\tau>\tau_{\mathrm{loop}}$, the fractional frequency instability
of the probe light stabilized to the atomic resonance is
$5-7\times 10^{-15}~\tau^{-1/2}$, in good agreement with the
quantum-limited instability expected for our experimental
conditions (trap secular frequencies, ion temperature and
measurement cycle time).

In order to compare the Hg$^+$ optical frequency standard to other
frequency standards, some fraction of the 563~nm light at the
output of AOM1 is launched into a 180~m long optical fiber and
delivered into a separate room where it is frequency shifted by
AOM3. AOM3 is used to actively cancel the optical path length
fluctuations of the fiber in order to preserve the high degree of
coherence of the light \cite{Young1999b,Ma1996}. 

The frequency of the 563~nm light is measured with an optical
frequency comb generated by a mode-locked Ti:Sapphire laser whose
femtosecond pulses are spectrally broadened in a microstructure
fiber \cite{Holzwarth2000,Diddams2000,Udem2001,Diddams2001}. The
optical spectrum at the output of the fiber consists of equally
spaced, phase-coherent modes with frequencies $f_n=n f_r+f_0$,
where $f_r$ is the repetition rate of the mode-locked laser, $n$
is an integer and $f_0$ is a frequency offset caused by the
difference between the phase and group velocities in the laser
cavity. In our setup, the repetition rate is typically $f_r\sim
1$~GHz and the spectrum spans more than one octave from $\sim
520$~nm to $1170$~nm. The repetition rate is detected directly
with a fast photodiode. The offset frequency $f_{0}$ is detected
by the self-referencing method, where the frequency-doubled
red part of the comb $2f_n=2n f_r+2f_0$ is heterodyned with the
blue part of the comb $f_{2n}=2n f_r+f_0$. We also detect the
beatnote $f_{b}$ between the 563~nm light from the Hg$^+$ optical
frequency standard and the closest mode of the optical frequency
comb $f_{m}=m f_r+f_0$ ($m$ is known from previous coarse
measurements of the Hg$^+$ $S-D$ transition frequency). In
practice, $f_{0}$ and $f_{b}$ are phase-locked to precision
rf sources by acting on the cavity length and the
pump power of the femtosecond laser, respectively. Finally, $f_r$
is measured by counting the low-frequency beatnote between $f_r$
and a third rf synthesizer. 

In order to perform absolute optical frequency measurements in terms
of the SI second, all synthesizers and frequency counters are
referenced to the $5$~MHz output of a hydrogen maser with a typical
frequency instability $\sim 2\times 10^{-13}$ at 1~s.  The maser
itself is part of an ensemble of 5 masers and 3 commercial cesium
clocks used to realize the local NIST atomic time scale, which is in
turn periodically calibrated using the NIST-F1 cesium fountain primary
standard \cite{Jefferts2002}, as well as international cesium
standards.  The frequency of the reference hydrogen maser is known
within $1.8$ parts in $10^{15}$ with respect to the ground-state
hyperfine splitting of $^{133}$Cs.  As shown in recent investigations
\cite{Udem1999b,Diddams2002}, the additional noise and inaccuracy from
the optical frequency comb itself is negligible at this level.  Using
this setup, it is thus possible to perform absolute measurements of
the frequency of the Hg$^+$ optical standard (together with the
frequency of each component of the optical comb) with a fractional
frequency uncertainty of $1.8\times 10^{-15}$.  Typically,
measurements are performed for 2 hours, leading to a statistical (type
A) fractional frequency uncertainty of 2.4 parts in $10^{15}$, which
corresponds to a $2.5$~Hz uncertainty on the frequency of the 282~nm
stabilized light.

The full evaluation of all systematic effects of the Hg$^+$ optical
standard is still under way.  At the present time, $10$~Hz is a
conservative upper bound for the total systematic (type B)
uncertainty, as shown by the following preliminary analysis of
systematic shifts.  The second-order Zeeman frequency shift of the
clock transition is given by $\delta\nu_{Z}= K_{Z}^{(2)}B^{2},~
K_{Z}^{(2)}=-18.925(28)$ kHz~mT$^{-2}$ \cite{Itano2000}.  The typical
field $B\approx 0.3$~mT corresponds to a second-order Zeeman frequency
shift $\delta\nu_{Z}\sim -1.7$~kHz.  Therefore, the $1.5\times
10^{-3}$ fractional uncertainty on coefficient $K_{Z}^{(2)}$ leads to
a $2.6$~Hz uncertainty on the clock frequency.  For the same value of
the bias magnetic field, the sensitivity of the clock frequency to
field fluctuations is $11$~Hz~$\mu$T$^{-1}$.  In our unshielded
environment, $B$-field fluctuations up to $\pm 0.2~\mu\mathrm{T}$ have
been observed, leading to a $2.2$~Hz uncertainty.  The most
troublesome frequency shift is the electric-quadrupole shift
$\delta\nu_{Q}$ due to the coupling between the atomic
electric-quadrupole moment in the upper $^2 D_{5/2}(F=2,m_{F}=0)$
clock state with electric field gradients due, for example, to stray
charges on the trap electrodes.  From our trap geometry and the bias
voltages applied to compensate for stray electric fields, it is
possible to calculate an upper bound for an electric field gradient at
the location of the ion and to estimate that $\delta\nu_{Q}<1$~Hz
\cite{Itano2000}.  In addition, two successive versions of the trap
electrodes (gold and molybdenum surfaces) have been used, leading to
no detectable change of the clock frequency.  Similarly, the
background pressure of helium (the only species remaining with
significant pressure at 4.2~K) has been changed by more than one order
of magnitude (estimated by measuring the ion heating rate due to
background collisions) without producing any detectable frequency
shift.  Empirically, this implies that the He background pressure
shift is smaller than $1$~Hz.  At a temperature of $300$~K, the
black-body radiation shift is $-0.08$~Hz, and it is considerably lower
in the $4.2$~K cryogenic environment \cite{Itano2000}.  At the Doppler
cooling limit of $1.7$~mK, the second-order Doppler shift due to
thermal motion is $-0.003$~Hz.  Finally, by suitably compensating the
stray static electric field, systematic shifts related to the trapping
oscillating electric field (second-order Doppler shift due to
micromotion, AC Stark shift) are made smaller than $0.1$~Hz
\cite{Berkeland1998a}.

\begin{figure}
\includegraphics[width=7cm]{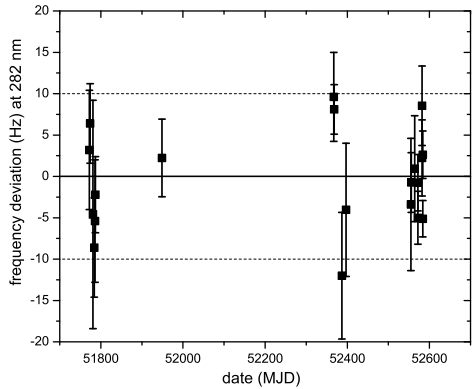}
\caption{\label{fig:measurements} Absolute frequency measurements
of the $^{199}$Hg$^{+}$ $^{2}S_{1/2} (F=0)\longleftrightarrow
~^{2}D_{5/2} (F=2)$ transition with respect to the $^{133}$Cs
ground state hyperfine splitting defining the SI second. The plot
shows the deviation of each measurement from the weighted average
value with its statistical $\pm 1\sigma$ error bar. The horizontal
axis is the Modified Julian Date ($\mathrm{MJD}~52582=$Nov. 4,
2002).}
\end{figure}
Over a period of two years, we have performed 20 measurements of
the $^{199}$Hg$^{+}$ $S-D$ transition frequency with respect to
the $^{133}$Cs ground state hyperfine splitting. Figure
\ref{fig:measurements} shows these measurements, corrected for the
second-order Zeeman frequency shift, with their statistical (type
A) error bars. The absolute frequency of the Hg$^+$ optical
standard is given by the weighted average of these data:
$\nu_{\mathrm{Hg}}=1~064~721~609~899~143.4$~Hz. The total
statistical uncertainty is only $1.0$~Hz. Measurements in Fig.
\ref{fig:measurements} clearly show a reproducibility better than
$10$~Hz at $1.06\times 10^{15}$~Hz, the most stringent comparison
of optical and microwave frequencies to date. The total
uncertainty of the measurement is dominated by the $10$~Hz
systematic uncertainty of the Hg$^{+}$ clock. With a $10^{-14}$
fractional uncertainty and a measurement period of two years, our
measurement constrains a possible fractional variation of
$\nu_{\mathrm{Cs}}/\nu_{\mathrm{Hg}}$ at the level of $\pm 7\times
10^{-15}$~yr$^{-1}$ ($1\sigma$ uncertainty).

This result can be interpreted as constraining a possible fractional
variation of $g_{\mathrm{Cs}}(m_e/m_p) \alpha^{6.0}$ at the same
level: $\pm 7\times 10^{-15}$~yr$^{-1}$.  Assuming that any change in
this quantity is due to the $\alpha^{6.0}$ factor, we derive an upper
bound for a possible linear variation of the fine-structure constant:
$|\dot{\alpha}/\alpha | <1.2\times 10^{-15}$~yr$^{-1}$, a factor of
$30$ improvement over \cite{Prestage1995}.  However, there may be a
significant change of $g_{\mathrm{Cs}}$ or $m_e/m_p$ due to possible
variation of the strength of the strong and electroweak interactions. 
In fact, recent theoretical work argues that within the framework of a
grand unified theory a fractional variation of $\alpha$ is necessarily
accompanied by a fractional variation of $m_e/m_p$ that is $\sim 38$
times larger \cite{Calmet2002}.  Fortunately, comparisons between two
optical clocks will test the stability of $\alpha$ alone
\cite{Dzuba1999a}.  One interesting possibility is to compare the
$^{199}$Hg$^{+}$ $S-D$ transition to the 657~nm
$^{1}S_{0}(m=0)\longleftrightarrow ~^{3}P_{1}(m=0)$ transition in
neutral $^{40}$Ca \cite{Udem2001,Helmcke2002}.  With an independent
constraint to the stability of $\alpha$, comparisons involving
hyperfine transitions, such as the present $\nu_{\mathrm{Hg}}$ (optical)
vs $\nu_{\mathrm{Cs}}$ comparison or the $\nu_{\mathrm{Rb}}$
(microwave) vs $\nu_{\mathrm{Cs}}$ comparison
\cite{Bize2001,Marion2002}, will test the stability of the strong and
electroweak interactions.  Reference \cite{Karshenboim2000b}
investigates in more detail possible  laboratory tests of
the stability of fundamental constants.

We thank Robert Windeler of OFS laboratories for providing the
microstructure fiber and Thomas Udem for his contributions in the
early measurements.  This work was partially supported by the Office
of Naval Research and through a cooperative research and development
agreement with Timing Solutions, Inc., Boulder, CO. This work of an
agency of the U.S. government is not subject to U.S. copyright.

\bibliography{bibliography}

\begin{thebibliography}{24}
\expandafter\ifx\csname
natexlab\endcsname\relax\def\natexlab#1{#1}\fi
\expandafter\ifx\csname bibnamefont\endcsname\relax
  \def\bibnamefont#1{#1}\fi
\expandafter\ifx\csname bibfnamefont\endcsname\relax
  \def\bibfnamefont#1{#1}\fi
\expandafter\ifx\csname citenamefont\endcsname\relax
  \def\citenamefont#1{#1}\fi
\expandafter\ifx\csname url\endcsname\relax
  \def\url#1{\texttt{#1}}\fi
\expandafter\ifx\csname
urlprefix\endcsname\relax\def\urlprefix{URL }\fi
\providecommand{\bibinfo}[2]{#2}
\providecommand{\eprint}[2][]{\url{#2}}

\bibitem[{\citenamefont{Damour et~al.}(2002)\citenamefont{Damour, Piazza, and
  Veneziano}}]{Damour2002}
\bibinfo{author}{\bibfnamefont{T.}~\bibnamefont{Damour}},
  \bibinfo{author}{\bibfnamefont{F.}~\bibnamefont{Piazza}}, \bibnamefont{and}
  \bibinfo{author}{\bibfnamefont{G.}~\bibnamefont{Veneziano}},
  \bibinfo{journal}{Phys. Rev. Lett} \textbf{\bibinfo{volume}{89}},
  \bibinfo{pages}{081601} (\bibinfo{year}{2002}).

\bibitem[{\citenamefont{Webb et~al.}(2001)\citenamefont{Webb, Murphy, Flambaum,
  Dzuba, Barrow, Churchill, Prochaska, and Wolfe}}]{Webb2001}
\bibinfo{author}{\bibfnamefont{J.}~\bibnamefont{Webb}}\textit{~et~al.},
  \bibinfo{journal}{Phys. Rev. Lett.} \textbf{\bibinfo{volume}{87}},
  \bibinfo{pages}{091301} (\bibinfo{year}{2001}).

\bibitem[{\citenamefont{Shlyakhter}(1976)}]{Shlyakhter1976}
\bibinfo{author}{\bibfnamefont{A.~I.} \bibnamefont{Shlyakhter}},
  \bibinfo{journal}{Nature} \textbf{\bibinfo{volume}{264}},
  \bibinfo{pages}{340} (\bibinfo{year}{1976}).

\bibitem[{\citenamefont{Damour and Dyson}(1996)}]{Damour1996}
\bibinfo{author}{\bibfnamefont{T.}~\bibnamefont{Damour}} \bibnamefont{and}
  \bibinfo{author}{\bibfnamefont{F.}~\bibnamefont{Dyson}},
  \bibinfo{journal}{Nucl. Phys. B} \textbf{\bibinfo{volume}{480}},
  \bibinfo{pages}{37} (\bibinfo{year}{1996}).

\bibitem[{\citenamefont{Prestage et~al.}(1995)\citenamefont{Prestage, Tjoelker,
  and Maleki}}]{Prestage1995}
\bibinfo{author}{\bibfnamefont{J.~D.}~\bibnamefont{Prestage}},
  \bibinfo{author}{\bibfnamefont{R.~L.}~\bibnamefont{Tjoelker}}, \bibnamefont{and}
  \bibinfo{author}{\bibfnamefont{L.}~\bibnamefont{Maleki}},
  \bibinfo{journal}{Phys. Rev. Lett.} \textbf{\bibinfo{volume}{74}},
  \bibinfo{pages}{3511} (\bibinfo{year}{1995}).

\bibitem[{\citenamefont{Dzuba et~al.}(1999)\citenamefont{Dzuba, Flambaum, and
  Webb}}]{Dzuba1999a}
\bibinfo{author}{\bibfnamefont{V.~A.}~\bibnamefont{Dzuba}},
  \bibinfo{author}{\bibfnamefont{V.~V.}~\bibnamefont{Flambaum}}, \bibnamefont{and}
  \bibinfo{author}{\bibfnamefont{J.~K.}~\bibnamefont{Webb}},
  \bibinfo{journal}{Phys. Rev. A} \textbf{\bibinfo{volume}{59}},
  \bibinfo{pages}{230} (\bibinfo{year}{1999}).

\bibitem[{\citenamefont{Young et~al.}(1999{\natexlab{a}})\citenamefont{Young,
  Cruz, Itano, and Bergquist}}]{Young1999}
\bibinfo{author}{\bibfnamefont{B.}~\bibnamefont{Young}}\textit{~et~al.},
  \bibinfo{journal}{Phys. Rev. Lett.} \textbf{\bibinfo{volume}{82}},
  \bibinfo{pages}{3799} (\bibinfo{year}{1999}{\natexlab{a}}).

\bibitem[{\citenamefont{Rafac et~al.}(2000)\citenamefont{Rafac, Young, Beall,
  Itano, Wineland, and Bergquist}}]{Rafac2000}
\bibinfo{author}{\bibfnamefont{R.~J.} \bibnamefont{Rafac}}\textit{~et~al.},
  \bibinfo{journal}{Phys. Rev. Lett.}
  \textbf{\bibinfo{volume}{85}}, \bibinfo{pages}{2462} (\bibinfo{year}{2000}).

\bibitem[{\citenamefont{Dehmelt}(1975)}]{Dehmelt1975}
\bibinfo{author}{\bibfnamefont{H.}~\bibnamefont{Dehmelt}},
  \bibinfo{journal}{Bull. Am. Phys. Soc.} \textbf{\bibinfo{volume}{20}},
  \bibinfo{pages}{60} (\bibinfo{year}{1975}).

\bibitem[{\citenamefont{Bergquist et~al.}(1987)\citenamefont{Bergquist, Itano,
  and Wineland}}]{Bergquist1987}
\bibinfo{author}{\bibfnamefont{J.~C.}~\bibnamefont{Bergquist}},
  \bibinfo{author}{\bibfnamefont{W.~M.}~\bibnamefont{Itano}}, \bibnamefont{and}
  \bibinfo{author}{\bibfnamefont{D.~J.}~\bibnamefont{Wineland}},
  \bibinfo{journal}{Phys. Rev. A} \textbf{\bibinfo{volume}{36}},
  \bibinfo{pages}{428} (\bibinfo{year}{1987}).

\bibitem[{\citenamefont{Itano et~al.}(1993)\citenamefont{Itano, Bergquist,
  Bollinger, Gilligan, Heinzen, Moore, Raizen, and Wineland}}]{Itano1993}
\bibinfo{author}{\bibfnamefont{W.}~\bibnamefont{Itano}}\textit{~et~al.},
  \bibinfo{journal}{Phys. Rev. A.} \textbf{\bibinfo{volume}{47}},
  \bibinfo{pages}{3554} (\bibinfo{year}{1993}).

\bibitem[{\citenamefont{Ma et~al.}(1996)\citenamefont{Ma, Jungner, Ye, and
  Hall}}]{Ma1996}
\bibinfo{author}{\bibfnamefont{L.-S.} \bibnamefont{Ma}}\textit{~et~al.},
  \bibinfo{journal}{Opt. Lett.} \textbf{\bibinfo{volume}{19}},
  \bibinfo{pages}{1777} (\bibinfo{year}{1996}).

\bibitem[{\citenamefont{Young et~al.}(1999{\natexlab{b}})\citenamefont{Young,
  Rafac, Beall, Cruz, Itano, Wineland, and Bergquist}}]{Young1999b}
\bibinfo{author}{\bibfnamefont{B.}~\bibnamefont{Young}}\textit{~et~al.},
  in \emph{\bibinfo{booktitle}{Proc. of the 14$^{th}$ Int. Conf. on Laser
  Spectroscopy}}, edited by
  \bibinfo{editor}{\bibfnamefont{R.}~\bibnamefont{Blatt}},
  \bibinfo{editor}{\bibfnamefont{J.}~\bibnamefont{Eschner}},
  \bibinfo{editor}{\bibfnamefont{D.}~\bibnamefont{Leibfried}},
  \bibnamefont{and}
  \bibinfo{editor}{\bibfnamefont{F.}~\bibnamefont{Schmidt-Kaler}}
  (\bibinfo{publisher}{World Scientific}, \bibinfo{year}{1999}{\natexlab{b}}).

\bibitem[{\citenamefont{Udem et~al.}(2001)\citenamefont{Udem, Diddams, Vogel,
  Oates, Curtis, Lee, Itano, Drullinger, Bergquist, and Hollberg}}]{Udem2001}
\bibinfo{author}{\bibfnamefont{T.}~\bibnamefont{Udem}}\textit{~et~al.},
  \bibinfo{journal}{Phys. Rev. Lett.} \textbf{\bibinfo{volume}{86}},
  \bibinfo{pages}{4996} (\bibinfo{year}{2001}).

\bibitem[{\citenamefont{Holzwarth et~al.}(2000)\citenamefont{Holzwarth, Udem,
  H{\"a}nsch, Knight, Wadsworth, and Russell}}]{Holzwarth2000}
\bibinfo{author}{\bibfnamefont{R.}~\bibnamefont{Holzwarth}}\textit{~et~al.},
  \bibinfo{journal}{Phys. Rev. Lett.} \textbf{\bibinfo{volume}{85}},
  \bibinfo{pages}{2264} (\bibinfo{year}{2000}).

\bibitem[{\citenamefont{Diddams et~al.}(2000)\citenamefont{Diddams, Jones, Ye,
  Cundiff, Hall, Ranka, Windeler, Holzwarth, Udem, and
  H\"{a}nsch}}]{Diddams2000}
\bibinfo{author}{\bibfnamefont{S.}~\bibnamefont{Diddams}}\textit{~et~al.},
  \bibinfo{journal}{Phys. Rev. Lett.} \textbf{\bibinfo{volume}{84}},
  \bibinfo{pages}{5102} (\bibinfo{year}{2000}).

\bibitem[{\citenamefont{Diddams et~al.}(2001)\citenamefont{Diddams, Udem,
  Bergquist, Curtis, Drullinger, Hollberg, Itano, Lee, Oates, Vogel
  et~al.}}]{Diddams2001}
\bibinfo{author}{\bibfnamefont{S.}~\bibnamefont{Diddams}}\textit{~et~al.},
  \bibinfo{journal}{Science}
  \textbf{\bibinfo{volume}{293}}, \bibinfo{pages}{825} (\bibinfo{year}{2001}).

\bibitem[{\citenamefont{Jefferts et~al.}(2002)\citenamefont{Jefferts, Shirley,
  Parker, Heavner, Meekhof, Nelson, Levi, Costanzo, DeMarchi, Drullinger
  et~al.}}]{Jefferts2002}
\bibinfo{author}{\bibfnamefont{S.}~\bibnamefont{Jefferts}}\textit{~et~al.},
  \bibinfo{journal}{Metrologia}
  \textbf{\bibinfo{volume}{39}}, \bibinfo{pages}{321} (\bibinfo{year}{2002}).

\bibitem[{\citenamefont{Udem et~al.}(1999)\citenamefont{Udem, Reichert,
  Holzwarth, and H\"{a}nsch}}]{Udem1999b}
\bibinfo{author}{\bibfnamefont{T.}~\bibnamefont{Udem}}\textit{~et~al.},
  \bibinfo{journal}{Opt. Lett.} \textbf{\bibinfo{volume}{24}},
  \bibinfo{pages}{881} (\bibinfo{year}{1999}).

\bibitem[{\citenamefont{Diddams et~al.}(2002)\citenamefont{Diddams, Hollberg,
  Ma, and Robertsson}}]{Diddams2002}
\bibinfo{author}{\bibfnamefont{S.}~\bibnamefont{Diddams}}\textit{~et~al.},
  \bibinfo{journal}{Opt. Lett.} \textbf{\bibinfo{volume}{27}},
  \bibinfo{pages}{58} (\bibinfo{year}{2002}).

\bibitem[{\citenamefont{Itano}(2000)}]{Itano2000}
\bibinfo{author}{\bibfnamefont{W.}~\bibnamefont{Itano}}, \bibinfo{journal}{J.
  Res. Natl. Inst. Stand. Technol.} \textbf{\bibinfo{volume}{105}},
  \bibinfo{pages}{829} (\bibinfo{year}{2000}).

\bibitem[{\citenamefont{Berkeland et~al.}(1998)\citenamefont{Berkeland, Miller,
  Bergquist, Itano, and Wineland}}]{Berkeland1998a}
\bibinfo{author}{\bibfnamefont{D.~J.} \bibnamefont{Berkeland}}\textit{~et~al.},
  \bibinfo{journal}{J. Appl. Phys.} \textbf{\bibinfo{volume}{83}},
  \bibinfo{pages}{5025} (\bibinfo{year}{1998}).

\bibitem[{\citenamefont{Calmet and Fritzsch}(2002)}]{Calmet2002}
\bibinfo{author}{\bibfnamefont{X.}~\bibnamefont{Calmet}} \bibnamefont{and}
  \bibinfo{author}{\bibfnamefont{H.}~\bibnamefont{Fritzsch}},
  \bibinfo{journal}{Eur. Phys. J. C} \textbf{\bibinfo{volume}{24}},
  \bibinfo{pages}{639} (\bibinfo{year}{2002}).

\bibitem[{\citenamefont{Helmcke et~al.}(2002)}]{Helmcke2002}
    \bibinfo{author}{\bibfnamefont{J.}~\bibnamefont{Helmcke}}\textit{~et~al.},
    in \emph{\bibinfo{booktitle}{Proc. of the 2002 CPEM Conference}};
    \bibinfo{journal}{IEEE Trans. Instrum. Meas.} (to be
    published).

\bibitem[{\citenamefont{Bize et~al.}(2001)}]{Bize2001}
    \bibinfo{author}{\bibfnamefont{S.}~\bibnamefont{Bize}}\textit{~et~al.},
    in \emph{\bibinfo{booktitle}{Proc. of the 6$^{th}$ Symposium on Frequency Standards and
    Metrology}}.
  (\bibinfo{publisher}{World Scientific},
  \bibinfo{year}{2001}{\natexlab{b}}), p. 53.
  
  
\bibitem[{\citenamefont{H. Marion et~al.}(2001)}]{Marion2002}
    \bibinfo{author}{\bibfnamefont{H.}~\bibnamefont{Marion}}\textit{~et~al.},
    Submitted to \prl (2002). 

\bibitem[{\citenamefont{Karshenbo\"{\i}m}(2000)}]{Karshenboim2000b}
    \bibinfo{author}{\bibfnamefont{S.~G.} \bibnamefont{Karshenbo\"{\i}m}},
  \bibinfo{journal}{Can. J. Phys.} \textbf{\bibinfo{volume}{47}},
  \bibinfo{pages}{639} (\bibinfo{year}{2000}).

\end{thebibliography}

\end{document}